\documentclass[prl,superscriptaddress,twocolumn]{revtex4}

\usepackage{amsfonts,amssymb,amsmath}
\usepackage[]{graphics,graphicx,epsfig}
\usepackage{amsthm}

\def\identity{\leavevmode\hbox{\small1\kern-3.8pt\normalsize1}}

\renewcommand{\epsilon}{\varepsilon}

\begin{document}


\title{Discrete Time Quantum Walk Approach to State Transfer}

\author{Pawe{\l} Kurzy\'nski}
\email{cqtpkk@nus.edu.sgl}
\affiliation{Centre for Quantum Technologies,
National University of Singapore, 3 Science Drive 2, 117543 Singapore,
Singapore}
\affiliation{Faculty of Physics, Adam Mickiewicz University, Umultowska
85, 61-614 Pozna\'{n}, Poland.}

\author{Antoni W{\'o}jcik}%
\affiliation{Faculty of Physics, Adam Mickiewicz University, Umultowska
85, 61-614 Pozna\'{n}, Poland.}

\date{\today}

\begin{abstract}

We show that a quantum state transfer, previously studied as a continuous time process in networks of interacting spins, can be achieved within the model of discrete time quantum walks with position dependent coin.  We argue that due to additional degrees of freedom, discrete time quantum walks allow to observe effects which cannot be observed in the corresponding continuous time case. First, we study a discrete time version of the engineered coupling protocol due to Christandl {\it et. al.} [Phys. Rev. Lett. {\bf 92}, 187902 (2004)] and then discuss the general idea of conversion between continuous time quantum walks and discrete time quantum walks.

\end{abstract}

\maketitle

\section{Introduction}

Lattice models appear in quantum theory in various contexts. Just like most theoretical models in physics, they were first used to describe the properties of natural materials, however nowadays we are reaching the stage when we are no longer bounded by nature, since our ability to engineer our own new systems having desired physical properties is developing very rapidly. This allows testing of abstract theoretical lattice models in the laboratory with a help of new extraordinary systems like graphene \cite{graphene}, cold atoms in optical lattices \cite{ol}, photons in arrays of linear optical elements \cite{linopt}, or compounds of thin layers of different materials known as heterostructures \cite{hetero}. The new field of quantum information, which strongly relies on the capability of engineering and manipulating of quantum systems, has also rekindled interest in lattice models. In particular, it has been shown that lattice models offer the possibility of universal quantum computation \cite{childs1,lovett}.

In this work we consider two related fields of research: quantum walks on graphs \cite{FG,Ahar} and quantum information transfer in spin systems \cite{Bose,Kay}. In particular we are interested in the discrete time quantum walk realization of a quantum state transfer on a spin chain with position dependent couplings \cite{christandl}. It is known that special cases of spin lattice models with one magnetic excitation can be interpreted as continuous time quantum walks \cite{FG}. Also, it has been shown that the relation between the two types of quantum walks is not trivial \cite{strauch1,childs2}, because discrete time quantum walks require an additional degree of freedom known as a coin. We show for the first time how one can convert the position dependence of couplings into the position dependence of coins. We also argue that discrete time quantum walks provide a more general framework for the description of quantum diffusion on regular lattices, since due to the extra degree of freedom one gains more control over the walk than is allowed in a standard continuous time scenario. Moreover, the evolution of discrete time quantum walks is much easier to simulate on a classical computer. Our approach is based on the relation between discrete time quantum walks and the Dirac-like equation to which the underlying quantum walk is transformed in the continuous limit of both time and space \cite{strauch2, bracken, kurzynski}.

\section{Basic concepts}

\subsection{State transfer in spin networks}

Consider a network for which every vertex corresponds to a spin $1/2$. We say that there is an edge between vertex $i$ and $j$ if the corresponding spins interact. Most authors choose to consider an $XX$ model of spin-spin interactions \cite{Bose,Kay}, we follow them and set the Hamiltonian of the network to be of the form
\begin{equation}\label{XX}
H=\sum_{\{i,j\}\in E}J_{ij}\left(\sigma_x^i\sigma_x^j+\sigma_y^i\sigma_y^j\right),
\end{equation}
where $E$ denotes the set of edges and $\sigma_{x}^{i}$ is a Pauli $X$ matrix for the $i$'th spin. The above Hamiltonian conserves the total spin number along $Z$ axis $\sum_i \sigma_z^i$, therefore one may restrict studies to a subspace with a fixed number of excitations, where by excitation we mean a spin pointing up along $Z$ axis. In particular, we are interested in an evolution restricted to a subspace with no excitations, which is invariant under the evolution generated by the Hamiltonian, and  to a subspace with only one excitation. In the second case, the single excitation can be considered as a single particle walking on an underlying network.  

Next, imagine that there are two marked spins, a sender and a receiver, and that the whole network is in a ground state, i.e. in a subspace with no excitations. Then, the sender prepares his spin in a superposition $\alpha|\uparrow\rangle+\beta|\downarrow\rangle$, so that the whole network is in the state
\begin{equation}
(\alpha|\uparrow\rangle+\beta|\downarrow\rangle)_s\otimes|\downarrow \downarrow \dots \downarrow\rangle_{net}\otimes|\downarrow\rangle_r.
\end{equation}
The goal is to employ the evolution generated by the Hamiltonian to transfer the quantum state from $s$ to $r$ in finite time $T$
\begin{equation}
|\downarrow\rangle_s\otimes|\downarrow \downarrow \dots \downarrow\rangle_{net}\otimes(\alpha|\uparrow\rangle+\beta|\downarrow\rangle)_r,
\end{equation}
which is equivalent to transporting the excitation from spin $s$ to spin $r$.

There are many protocols which allow for a perfect, or a nearly perfect state transfer over spin chains. The basic techniques are: engineered couplings, wave packet encoding, and active control (see \cite{Bose,Kay}). Here we concentrate on engineered couplings. In particular, we will consider two protocols, namely the protocol of Christandl {\it et. al.} \cite{christandl} and a weakly coupled spin protocol \cite{wojcik1,wojcik2}.

\subsection{Continuous time quantum walks}

The state in continuous time quantum walks (CTQW) \cite{FG} is described by the position of a particle on a graph. The continuous evolution, which is governed by the Schr\"odinger equation, is determined by the Hamiltonian $H$ which is an $n \times n$ Hermitian matrix proportional to the adjacency matrix of the
underlying graph $G$ which has $n$ vertices
\begin{equation} \label{e04}
 H_{ij}\left\{
        \begin{array}{ll}
          \neq 0, & \hbox{iff $(i,j)\in E(G)$;} \\
          =0, & \hbox{else,}
        \end{array}
      \right.
\end{equation}
where $E(G)$, as before, denotes the set of edges of $G$. The probability that at time $t$ the walker, whose state is $|\psi(t)\rangle = e^{-iHt}|\psi(0)\rangle$, is located at vertex $x$ is $P_x(t)=|\langle x |\psi(t)\rangle|^2$, where $\{|x\rangle\}$ is the orthonormal basis spanning the space of vertices.

For CTQW an a graph all of  whose edges have equal weight, the corresponding Hamiltonian is simply the Laplacian of the underlying graph $L=D-A$, which is a diagonal matrix $D$ minus the graph adjacency matrix $A$ (whose entries $A_{ij}$ are either zero or one if vertices $i$ and $j$ are disconnected or connected, respectively). The diagonal entries $D_{jj}=d_j$ denote the degree of vertex $j$. As a result, the sum of all elements in each column and each row of the Laplacian is zero. When the Laplacian plays the role of the Hamiltonian, the diagonal entries are important only in the case of irregular graphs. For regular graphs they are all equal, and since one can freely add or subtract from the Hamiltonian multiples of identity without changing the dynamics of the system, they can be neglected. In case of a general CTQW on a chain the only nonzero elements of the Hamiltonian are $H_{j,j+1}$, $H_{j,j-1}$ and $H_{jj}$.  In case of a single excitation subspace of a spin network, the elements of the corresponding CTQW Hamiltonian are given exactly by the strengths of spin couplings $H_{ij}=J_{ij}$.

\subsection{Discrete time quantum walks}

In discrete time quantum walks (DTQW) \cite{Ahar} the state of the system $|x,c\rangle$ is described by the position of a walker on a graph $x$ and by the state of an auxiliary system $c$ which determines the direction the walker is going to take in the next step. This auxiliary system is often referred to as the coin, which for a walk on a one-dimensional graph like a chain or cycle is simply a two-level system $c=\leftarrow,\rightarrow$. In such cases, one step of the evolution is given by a unitary operator $U=SC$, which is a product of the coin operator
\begin{eqnarray}
C|x,\rightarrow\rangle&=& \cos\theta|x,\rightarrow\rangle-\sin\theta|x,\leftarrow\rangle, \label{e1}\\
C|x,\leftarrow\rangle&=&\sin\theta|x,\rightarrow\rangle+\cos\theta|x,\leftarrow\rangle, \label{e2} 
\end{eqnarray}
and the conditional shift operator
\begin{eqnarray}
S|x,\rightarrow\rangle&=&|x+1,\rightarrow\rangle , \label{e3}\\
S|x,\leftarrow\rangle&=&|x-1,\leftarrow\rangle. \label{e4} 
\end{eqnarray}Therefore one step is given by
\begin{eqnarray}
U|x,\rightarrow\rangle&=& \cos\theta|x+1,\rightarrow\rangle-\sin\theta|x-1,\leftarrow\rangle, \label{e5}\\
U|x,\leftarrow\rangle&=&\sin\theta|x+1,\rightarrow\rangle+\cos\theta|x-1,\leftarrow\rangle. \label{e6} 
\end{eqnarray}
The parameter $\theta$ denotes the coin flip rate. Equivalently, the unitary evolution operator of one step can be written as
\begin{equation} 
U=e^{i p \sigma_z}e^{i\sigma_y \theta}, \label{e03}
\end{equation} 
where $p$ is the momentum operator and $\sigma_j$ are Pauli matrices acting on the coin space. 

\begin{figure}
\scalebox{1.0}  {\includegraphics[width=8truecm]{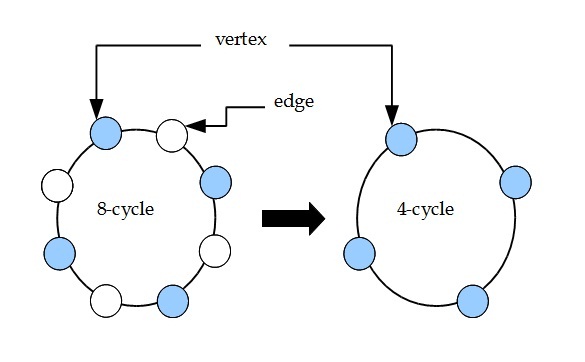}}
\vspace{-0.5cm}
\caption{\label{f1} Due to the fact that interference phenomenon occurs for the second nearest neighbors even vertices can be interpreted as edges.}
\end{figure}

\begin{figure}
\scalebox{1.0}  {\includegraphics[width=8truecm]{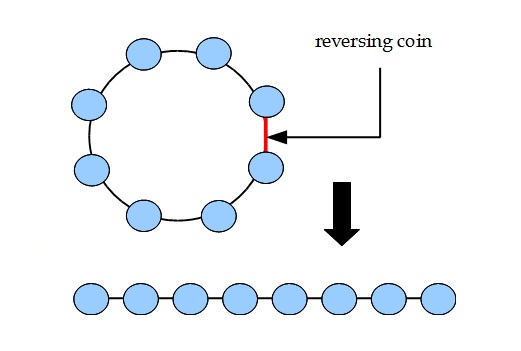}}
\vspace{-0.5cm}
\caption{\label{f2} DTQW on a finite chain is effectively simulated by a walk on a cycle with one reversing edge.}
\end{figure}

In DTQW on chains and even cycles, interference occurs only between positions separated by a distance of two (second nearest neighbors), therefore one can consider two independent walks: one on even vertices and another on odd vertices. Due to this fact even vertices can be treated as {\it edges} as in Fig. \ref{f1}, which will be crucial for our work.
Since we are interested in the walk on vertices, not on {\it edges}, in our case one step of evolution will be given by $U^2$ and the initial state will be always supported on the vertex space only, as in the Ref. \cite{strauch1}. Moreover, throughout this work we will consider DTQW on finite chains, which can be simulated by walks on cycles with a reversing coin at one vertex, i.e. a coin for which coin flip rate $\theta=\pi/2$ --- see Eqs. (\ref{e1},\ref{e2}) and Fig. \ref{f2}.

\section{Continuous limit of quantum walks}

The continuous limit of DTQW on a chain results in the one-dimensional Dirac equation \cite{strauch2,bracken,kurzynski}. The operator (\ref{e03}) can be rewritten as
\begin{equation} 
U=U(t)=e^{i p \sigma_z v t}e^{i\sigma_y \omega t}, \label{e7}
\end{equation} 
where $v$ is the velocity and $\omega$ is the angular frequency. The two parameters are chosen so that for a unit time step $vt=1$ and $\omega t=\theta$, thus recovering Eq. (\ref{e03}). Taking infinitesimal time steps and applying the Trotter theorem one obtains
\begin{equation} 
\lim_{n\rightarrow \infty}U\left(t/n\right)^{n}=e^{i (p \sigma_z v + \sigma_y \omega) t}. \label{e8}
\end{equation}
The term in the bracket above corresponds to the Hamiltonian of the system, which has exactly the form of the one-dimensional Dirac Hamiltonian
\begin{equation} 
H=cp\sigma_z + mc^2 \sigma_y. \label{e9}
\end{equation}
The correspondence between (\ref{e8}) and (\ref{e9}) is fully established when one identifies $v$ with the speed of light and $\omega$ with the mass of the particle. In particular, $\omega t =\theta$, however we take $t=1$, therefore the mass of the quantum walker corresponds to the coin flip rate $\theta$. However, due to the discrete nature of time and the periodicity of coin operator $e^{i\sigma_y\theta}$, one is unable to differentiate between $\theta$ and $\theta+k 2\pi$. On the other hand, in the limit of infinite mass the particle should be almost immobile and the relation between the Dirac equation and DTQW requires $\theta \rightarrow \frac{\pi}{2}$, hence one often takes $m=\tan\theta$ (see \cite{strauch2}). Note that for small angles $\tan\theta\approx \theta$. It is also important to notice, that due to the structure of the conditional translation operator the continuization of time is intrinsically combined with the continuization of space, i.e. either both space and time are continuous, or both are  discrete. 

The Hamiltonian of CTQW on a chain with uniform couplings is the following: $H_{i,i+1}=H_{i+1,i}=-J$ and $H_{i,i}=2J$, however for regular graphs diagonal elements can be neglected and put equal to zero. The action of the Hamiltonian on a state localized at $x$ is
\begin{equation} 
H|x\rangle=-J\left(|x+1\rangle-2|x\rangle+|x-1\rangle\right). \label{e10}
\end{equation}
The above is a discrete version of the Laplacian. The continuization of the above leads to
\begin{equation} \label{lap}
H=-J\frac{d^2 }{d x^2}. 
\end{equation}
Obviously, this Hamiltonian describes a free particle in one dimension with $J=\frac{1}{2m}$, therefore the coupling constant $J$ can be interpreted as an inverse of the walker's mass.

In the continuous limit, the quantum walker can be considered as a relativistic (DTQW), or as a non-relativistic free particle (CTQW). Main parameters governing the dynamics of the walk, namely coins and couplings, are related to the particle's mass. Strauch \cite{strauch1} showed that continuous quantum walk can be obtained from discrete one in the limit $\theta \rightarrow \frac{\pi}{2}$. Heuristically, this limit corresponds to a very heavy particle, for which it is much harder to observe relativistic effects, therefore it corresponds to a relativistic to non-relativistic transition. Another important feature of quantum walks is that in both models space is discrete, but only for DTQW time is also discrete. Once again, a heuristic explanation of this fact can be given as follows. Let the two quantum walks be discrete versions of Dirac and Schr\"odinger free particle, respectively. In non-relativistic model time and space are treated separately, therefore there is nothing strange in discretizing only one of them. However, in relativistic physics time and space are combined and one has to discretize the whole spacetime at once, since it would be problematic to define consistent Lorentz transformations on spacetime which is only partially discrete.

\section{Perfect State Transfer in DTQW}

Let us apply the analogy between mass, coins and couplings to show that a perfect state transfer from one position to another can be realized within DTQW. We concentrate on the acclaimed perfect state transfer protocol introduced by Christandl {\it et. al.} \cite{christandl}. This protocol relies on properly engineered couplings, i.e. properly chosen terms of the Hamiltonian governing the corresponding CTQW. These couplings depend on position, therefore by our analogy, the corresponding particle has position dependent mass. On the other hand, mass is related to coin flip rate and so the corresponding DTQW should have position dependent coin. Below, we examine how the choice of such coins affects the dynamics of DTQW.

The chain coupling between nodes $n$ and $n+1$ in the protocol is given by $J_n=\frac{\lambda}{2}\sqrt{n(N-n)}$, where $n=1,2,\dots,N-1$ and $\lambda$ is a real parameter common for all couplings. As a result, the effective mass of the particle for the transfer between positions $n$ and $n+1$ is 
\begin{equation}\label{mass}
m_n=\frac{1}{2J_n}=\frac{1}{\lambda\sqrt{n(N-n)}}.
\end{equation}
Next, consider a step operator of the form
\begin{equation} \label{e13}
U=U(n)=e^{ip\sigma_z}e^{i\sigma_y\theta_n},
\end{equation}
where $\theta_n$ depends on position. The matrix form of the coin operator
is given by
\begin{equation} \label{pdc}
C_n=\left(
       \begin{array}{cc}
         \cos \theta_n & \sin \theta_n \\
         -\sin \theta_n & \cos \theta_n \\
       \end{array}
     \right).
\end{equation}
Recall that we consider a double step operator $U^2$, since our walk on a chain of length $N/2$ is simulated by a walk on an even N-cycle for which even positions correspond to edges and we are interested in a walk on positions corresponding to vertices. Coins associated with edge positions correspond to mass, but there are also coins associated with vertices --- odd positions of the cycle. Here, we set all of them to be identities ($\theta_{2k-1}=0$), although we note that a different choice (equal for all vertices) leads to similar results. The edge coins (even positions) are chosen according to the analogy between couplings and mass ($m=\tan\theta$)
\begin{equation} \label{theta}
\theta_{2k}=\arctan\frac{1}{2\lambda\sqrt{k\left(\frac{N}{2}-k\right)}},
\end{equation}
where $k=1,2,\dots,N/2-1$. 

Our studies show that the transfer strongly depends on parameter $\lambda$. For small values of $\lambda$ (large mass limit) the transfer is perfect. Around $\lambda=\frac{\pi}{N}$ it starts to drop down and it recovers back to perfect transfer for $\lambda$ of the order $O(1)$ and greater (small mass limit). We show that the nature of transport for the two limiting cases is drastically different. In the small mass limit perfect transfer occurs due to dispersion-less nature of the operator $e^{ip\sigma_z}$ whose spectrum is linear, whereas in the large mass limit the generator of the operator $U^2$ is effectively given by a Hermitian operator whose action is the same as the action of the Hamiltonian introduced by Christandl {\it et. al.} \cite{christandl}. Interestingly, in both cases we found that the transfer fidelity does not depend on the initial coin state. Moreover, we checked numerically that a number of initial coin states, among which were eigenvectors of the three Pauli matrices, are always perfectly transfered. This fact allows us to conjecture that in DTQW case not only particle, but also its intrinsic coin state is perfectly transfered: $\alpha|1,\rightarrow\rangle+\beta|1,\leftarrow\rangle~ \rightarrow~ \alpha|N-1,\rightarrow\rangle+\beta|N-1,\leftarrow\rangle$.

\begin{figure}
\scalebox{1.0}  {\includegraphics[width=8truecm]{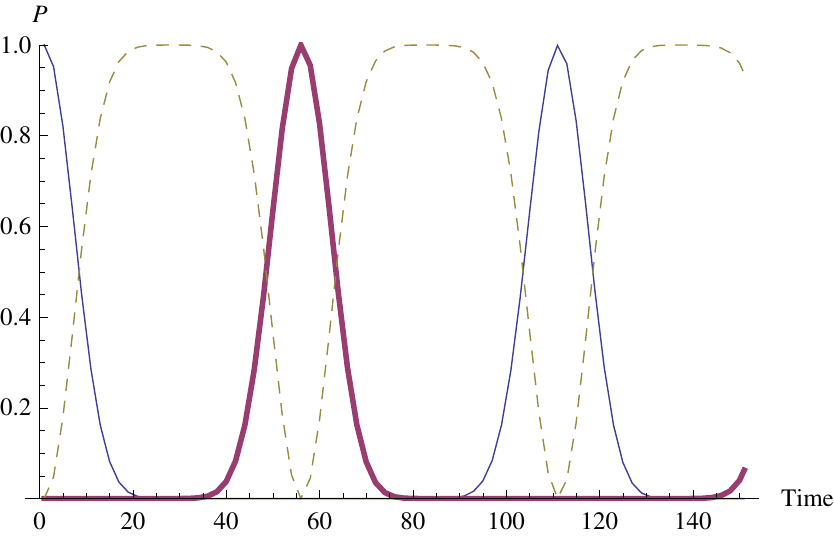}}
\caption{\label{f3} Perfect state transfer from position $1$ to position $N-1$ for DTQW with position dependent coin ($N=30$ and $\lambda=0.03$). Thin blue line --- probability of occupying position 1; thick purple line --- pobability of occupying position $N-1$; dashed line --- probability of occupying remaining vertices of the chain. }
\end{figure}
In Figure \ref{f3} we present the transfer of particle from position $1$ to position $N-1$ for the corresponding DTQW with respect to evolution operator $U^2$. In this particular case $N=30$ and $\lambda=0.03$. As one can see transfer is perfect, and moreover the dynamics is periodic because after twice the transfer time the system goes back to the initial state, as in the continuous time case.

\begin{figure}
\scalebox{1.0}  {\includegraphics[width=8truecm]{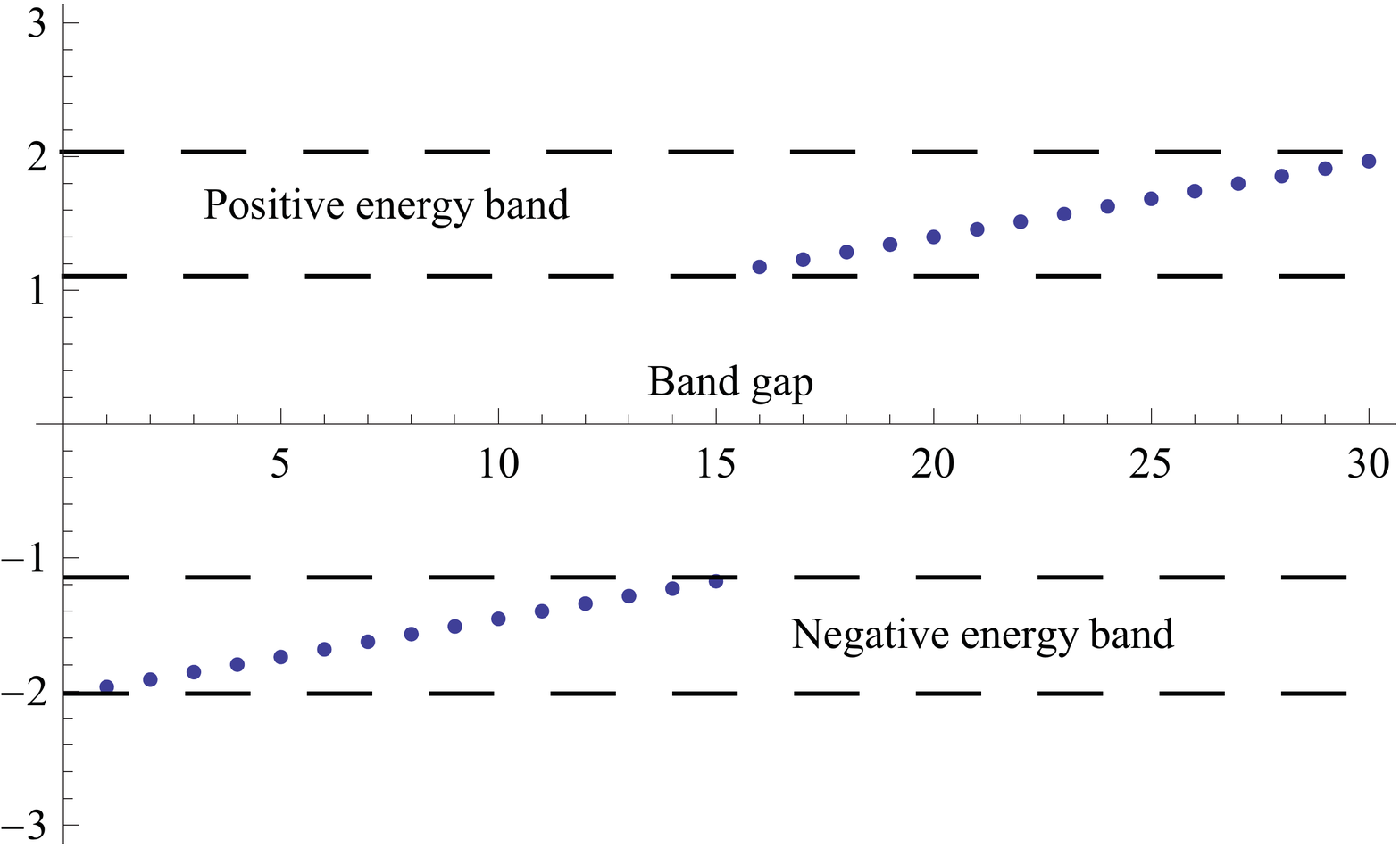}}
\caption{\label{f4} Arguments $\varphi$ of eigenvalues $e^{i\varphi}$ of a position dependent coin DTQW double step operator $U^2$  for $N=30$ and $\lambda=0.03$. Each dot corresponds to a doubly degenerated eigenvalue.}
\end{figure}

The main idea of the authors of the protocol \cite{christandl} was to find a mirror symmetric Hamiltonian with harmonic spectrum, which would provide periodicity of evolution and a perfect transfer between mirror symmetric positions for times equal to half of a period. Numerical simulations confirm that the spectrum of $U^2$ is also harmonic, yet it posses an additional relativistic property (see Fig. \ref{f4}). 
The arguments of eigenvalues of $U^2$ can be interpreted as quasi-energies. They split into two bands, which can be related to positive and negative energy bands just like in the Dirac equation. Moreover, the two bands are separated by a band gap corresponding to the rest energy $2mc^2$. Due to the discrete nature of time for quasi-energies the band gap occurs twice, the first centered around zero and the second centered around $\pi$. Numerical simulations show that the width of the band gap is related to the inverse of $\lambda$: in the limit $\lambda\rightarrow 0$ the band gap goes to $\pi$, and for $\lambda\rightarrow \infty$ the band gap goes to zero. This is as expected, since the corresponding mass scales as $\frac{1}{\lambda}$ --- see Eq. (\ref{mass}).

\subsection{Large mass limit}

Recall the formulae for position dependent coin (\ref{pdc}) and coin flip rate (\ref{theta}) and recall that $\arctan\theta$  can be expressed as
\begin{equation}\label{arcus}
\arctan\theta=\arcsin\left(\frac{\theta}{\sqrt{1+\theta^2}}\right)=\arccos\left(\frac{1}{\sqrt{1+\theta^2}}\right).
\end{equation}
The entries of the coin matrix can be rewritten as
\begin{equation}\label{coinarc}
\cos\theta_{2k}=\frac{2\lambda\gamma_k}{\sqrt{1+4\lambda^2\gamma_k^2}},~~\sin\theta_{2k}=\frac{1}{\sqrt{1+4\lambda^2\gamma_k^2}},
\end{equation}
where $\gamma_k=\sqrt{k\left(\frac{N}{2}-k\right)}$. In the large mass limit $\lambda\rightarrow 0$, which allows us to keep only the highest terms of Taylor expansion
\begin{equation}\label{coinarc2}
\cos\theta_{2k}\approx 2\lambda \gamma_k,~~\sin\theta_{2k}\approx 1.
\end{equation}
The above limit resembles the one of Strauch \cite{strauch1} $\theta_{2k} \rightarrow \frac{\pi}{2}$. The action of the double step operator $U^2$ on positions corresponding to vertices (odd positions) can be approximated by
\begin{eqnarray}
U^{2}|2k-1,\rightarrow\rangle & \approx & 2\lambda\gamma_k|2k+1,\rightarrow\rangle+|2k-1,\leftarrow\rangle,  \\
U^{2}|2k-1,\leftarrow\rangle & \approx & 2\lambda\gamma_{k-1}|2k-3,\leftarrow\rangle+|2k-1,\rightarrow\rangle.
\end{eqnarray}
Let us also introduce the following {\it almost normalized} states
\begin{eqnarray}
|\psi_{\pm}(k)\rangle=\frac{1}{\sqrt{2}}\left(|2k-1,\rightarrow\rangle \pm i |2k-1,\leftarrow\rangle \right. \nonumber \\
\left. \pm i 2\lambda\gamma_k|2k+1,\rightarrow\rangle- 2\lambda\gamma_{k-1}|2k-3,\leftarrow\rangle \right),
\end{eqnarray}
and normalized states
\begin{equation}
|\phi_{\pm}(k)\rangle=\frac{1}{\sqrt{2}}\left(|2k-1,\rightarrow\rangle \pm i |2k-1,\leftarrow\rangle \right).
\end{equation}
Note that $|\phi_{\pm}(k)\rangle$ and $|\psi_{\pm}(k)\rangle$ are almost equal up to the factor of $O(\lambda)$.
The action of double step operator on $|\psi_{\pm}(k)\rangle$ yields
\begin{eqnarray}
U^2|\psi_{\pm}(k)\rangle &=& i\left(|\phi_{\pm}(k)\rangle-i2\lambda \gamma_k |\phi_{\pm}(k+1)\rangle\right. \nonumber \\
 &-& \left.  i2 \lambda \gamma_{k-1} |\phi_{\pm}(k-1)\rangle \right) + O(\lambda^2)
\end{eqnarray}
Now, it is enough to recognize that $2\lambda\gamma_k=J_{k,k+1}$ from the protocol \cite{christandl} and that for short times the action of the unitary operator can be approximated by $U\approx \openone-iHt$, where $H$ is the Hamiltonian, therefore
\begin{equation}
U^2|\psi_{\pm}(k)\rangle \approx i(\openone - iH)|\phi_{\pm}(k)\rangle.
\end{equation}
The above Hamiltonian is exactly the Hamiltonian of Christandl {\it et. al.}, therefore the CQTW behavior is restored.  Moreover, the action of $H$ does not mix $\pm$ states, which is the reason why we observe not only the transfer of the walker, but also the transfer of its intrinsic coin state.

\subsection{Small mass limit}

Let us consider $\lambda\rightarrow \infty$. In this case a dispersion-less movement dominates the evolution, since Eq. (\ref{coinarc}) is approximately
\begin{equation}
\cos\theta_{2k}\approx 1,~~\sin\theta_{2k}\approx 0,
\end{equation}
therefore the double step operator simplifies to $U^2=e^{-2ip\sigma_z}$, except at position $N$, where we apply the reversing coin which also causes a $\pi$ phase shift of the particle coming from the left. The transfer occurs in $N/2$ double steps. The initial state $\alpha|1,\rightarrow\rangle+\beta|1,\leftarrow\rangle$ evolves in the following way. The first part of this state, which is the right moving part, moves to the right, is reflected at position $N$ and goes back to position $N-1$ to finish in state $-\alpha|N-1,\leftarrow\rangle$. The second left moving part goes left to $0\equiv N \mod N$, gets reflected and moves right to position $N-1$ to finish in state $\beta|N-1,\rightarrow\rangle$. This also works for any initial state localized at $x$, for which the destination point is $N-x$. In order to restore the final coin state to the initial state one has to apply the $\sigma_y$ operation.  

\subsection{Transition between the two types of transfer}

The unitary evolution operator $U$ is the product of two operators $S$ and $C$. Transfer properties rely on the interplay between this pair. In the small mass limit $C$ does not play any significant role, since it is close to identity. On the other hand, in the large mass limit $C$ is crucial, because it makes the  evolution similar to the one of the protocol \cite{christandl}. As we pointed out in the beginning of this section, we observed that the transition between the two types of behavior occurs around $\lambda=\frac{\pi}{N}$.  Here, we give the heuristic explanation for this value. 

The time of transfer over a chain of length $N$ for the protocol \cite{christandl} is $T=\frac{\pi}{\lambda}$. In our case we consider chains which are effectively of length $N/2$, that is why we have to multiply $\lambda$ by two, which is evident in Eq. (\ref{theta}). As a result, the corresponding time of transfer should be given by ${\cal T}=\frac{\pi}{2\lambda}$ and indeed this is approximately what we observe. Next, let us recall that DTQW is related to relativistic quantum mechanics via its similarity to the Dirac equation. In relativistic physics a distance traveled by a particle in time $t$ can be at most $ct$.  Similarly, in DTQW in one step the particle can move only to neighboring positions, that is why in order to travel the distance from one end of the chain to another the particle needs to take at least $N/2$ steps. This gives the critical value of $\lambda_{cr}=\frac{\pi}{N}$ above which the particle would move with velocity greater than the {\it speed of light}. Since this is impossible, a relativistic behavior has to dominate the evolution and as a consequence the transition between the two types of behavior can be interpreted as a non-relativistic to relativistic transition. 

\subsection{Other techniques for perfect transfer}

As mentioned earlier, nearly perfect transfer in spin networks can be obtained in many ways. In principle, all transfer techniques can be employed in DTQW. For example, wave packed encoding (see \cite{Bose} and references therein) is straightforward since one can prepare the same spatial wave packet in DTQW as the one considered in CTQW. However, position dependent coupling method is the one for which the correspondence between DTQW and CTQW is the most nontrivial. Bellow we would like to mention another example based on our coin-mass-coupling analogy. We present DTQW version of a protocol in which two spins are weakly coupled to ends of a spin chain with uniform couplings \cite{wojcik1,wojcik2}.

The CTQW Hamiltonian of a chain with weakly coupled ends is determined by coupling constants $J_{i,i+1}=J$ for $i=2,\dots,N-2$ and $J_{1,2}=J_{N-1,N}=aJ$, where $a \ll1$. Due to weak coupling to the rest of the chain an interaction of the first and the last spin with the middle part can be treated as a perturbation. Without interaction the two spins would be in a degenerate state. Weak coupling breaks the degeneracy and perturbation theory predicts that there are new eigenstates which are even superpositions of two states: a state in which the first spin is up and one in which the last spin is up, or three states: the first spin up, the last spin up, or the middle of the chain being in one of its unperturbed eigenstates. The second case happens when the degenerated eigenvalue of the two spins  is the same as one of the eigenvalues of the chain. Nearly perfect transfer is possible because initial and final states are almost completely supported on these two/three states only (for the detailed discussion see \cite{wojcik1,wojcik2}).  

In DTQW the above protocol can be realized in the following way. As before, we can simulate $N/2$-chain on  $N$-cycle with one reversing edge. All coins are the same $\theta_k=\theta$, except $\theta_2=\theta_{N-2}=\frac{\pi}{2}-\varepsilon$. We observe that DTQW behaves exactly like the corresponding CTQW. Moreover, we also observe the transfer of the coin state.
\begin{figure}
\scalebox{1.0}  {\includegraphics[width=4truecm]{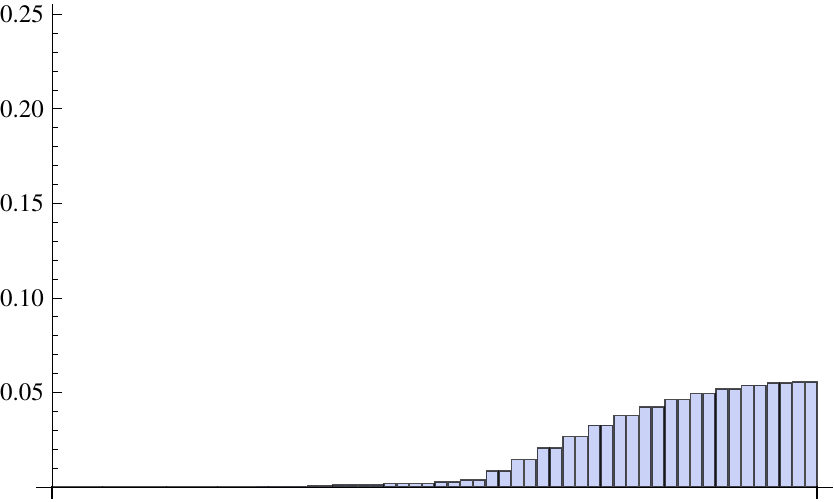}} {\includegraphics[width=4truecm]{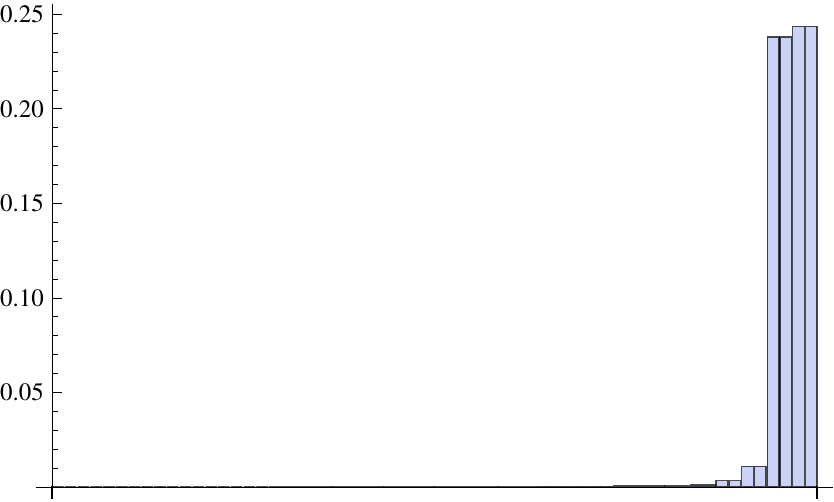}}
\caption{\label{f5} Eigenstate population for $N=30$, $\theta=\pi/4$ and initial state $|1,\rightarrow\rangle$. Left: $\theta_1=\theta_{N-1}=\theta=\pi/4$. Right: $\theta_2=\theta_{N-2}=\pi/2 - \varepsilon$, where $\varepsilon=\frac{\pi}{2N}$; the initial state is almost entirely supported on four eigenstates only.}
\end{figure}
\begin{figure}
\scalebox{1.0}  {\includegraphics[width=8truecm]{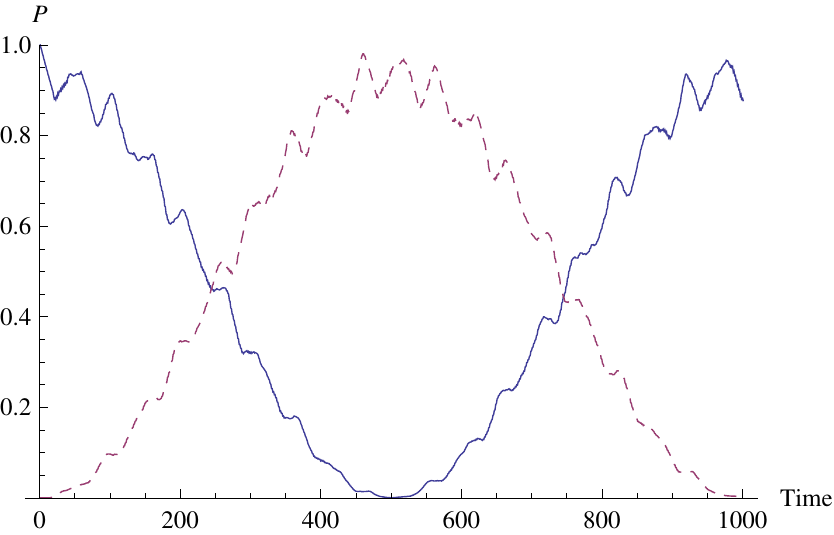}}
\caption{\label{f6} Probability of population of positions $1$ (solid) and $N-1$ (dashed) in time. $N=30$, $\theta=\frac{\pi}{4}$ and $\varepsilon=\frac{\pi}{2N}$.}
\end{figure}
 In Fig. \ref{f5} we show the eigenstate population for $N=30$, $\theta=\frac{\pi}{4}$ and initial state $|1,\rightarrow\rangle$. In case $\varepsilon \ll 1$ the initial state is almost entirely supported on four eigenstates (two positive energy eigenstates and two negative energy eigenstates). In Fig. \ref{f6} we present the population of positions $1$ and $N-1$ in time.  The time of transfer scales as $O\left(\frac{1}{\varepsilon^2}\right)$.  

We also observe another interesting phenomenon. If one changes the coin operator at position $2$ from $e^{i\sigma_y \frac{\theta_2}{2}}$ to  $e^{i\sigma_x \frac{\theta_2}{2}}$, the transfer is suppressed and the state is essentially localized at position $1$ for all time. The reason for this behavior is that the initial energy at position $1$ differs from the energy the particle would have at position $N-1$, therefore the transport is forbidden due to energy conservation. In the next section we will explain this in further detail. 

\section{How to convert between CTQW and DTQW?}

The conversion of CTQW into DTQW was considered by Childs \cite{childs2}, however his approach was purely mathematical and did not take into account the physical properties of the underlying quantum walks. On the other hand, our approach is based on physical aspects of quantum walks, and hence by studying the DTQW version of a continuous process one can learn many facts about the physics governing the corresponding system. In this section we show how to, in general, convert an arbitrary CTQW on a chain into DTQW. Once again, we are going to implement walk on a $N/2$-chain using $N$-cycle with vertex and edge positions.  

Recall that in general CTQW on a chain the only nonzero elements of the Hamiltonian are $H_{j,j+1}$, $H_{j,j-1}$ and $H_{jj}$. Diagonal terms, which are necessarily real, correspond to the potential energy  
\begin{equation}\label{pot}
u(j)=H_{jj}. 
\end{equation}
Potential energy has been already considered in DTQW, mostly in the context of localization (see for example \cite{wojcik3}). The kinetic energy for the transition between positions $j$ and $j+1$, which is proportional to the second power of momentum divided by the mass, corresponds to the real part of the off-diagonal terms $H_{j,j+1}$ and $H_{j+1,j}$. In this case $\text{Re}[H_{j,j+1}]=\frac{1}{2m_{j,j+1}}$. It was studied in the previous sections. The mass can be expressed as
\begin{equation}\label{m}
m_{j,j+1}=\frac{1}{2\text{Re}[H_{j,j+1}]}.
\end{equation}

Finally, let us consider the imaginary part of $H_{j,j+1}$. First of all, note that operator proportional to $-i|j\rangle\langle j+1|+i|j+1\rangle\langle j|$, which gives rise to nonzero imaginary part of $H_{j,j+1}$, can be approximated as a first power of momentum operator for positions $j$ and $j+1$ (see \cite{childs3}). Indeed,
\begin{eqnarray}
\left(\sum_{j}-i|j\rangle\langle j+1| +i|j+1\rangle\langle j|\right)^2 = \nonumber \\ 
\sum_j  2|j\rangle\langle j|-|j+2\rangle\langle j|-|j\rangle\langle j+2|.
\end{eqnarray}
The above operator resembles the Laplacian, which is the kinetic energy operator proportional to the square of momentum. The first power of momentum can appear in the Hamiltonian if there is a vector potential $\vec{A}$ acting on the particle. In this case we have to substitute $(\vec{p})^2\rightarrow \left(\vec{p}-\frac{e}{c}\vec{A}\right)^2$, where $e$ is the charge of the particle. As before, in the following, we assume $c=1$ and we also set $e=1$. Using this analogy one arrives at 
\begin{equation}
\text{Im}[H_{j,j+1}]=\frac{A_{j,j+1}}{m_{j,j+1}},
\end{equation} 
where in general both, the vector potential and the mass, are position dependent. The vector potential alone can be estimated as follows
\begin{equation}\label{vecpot}
A_{j,j+1}=\frac{1}{2}\frac{\text{Im}[H_{j,j+1}]}{\text{Re}[H_{j,j+1}]}.
\end{equation}

Our approach is based on the analogy between DTQW and the Dirac equation, whose general time-independent  one-dimensional form is
\begin{equation}\label{dirac}
\left(-i\frac{\partial}{\partial x} - A(x)\right)\sigma_z \psi+ m(x)\sigma_y \psi = \left(i\frac{\partial}{\partial t} - u(x)\right)\psi.
\end{equation} 
At this point we can go to DTQW via application of the Trotter formula. One step of the corresponding DTQW is given by
\begin{equation}\label{genqw}
U=SCV=e^{ip\sigma_z}e^{i\left(m(j)\sigma_y-A(j)\sigma_z\right)}e^{i u(j)},
\end{equation} 
where $j$ denotes position. To include the account of the scalar potential, we introduced a new operator $V=e^{iu(j)}$. Moreover, the new coin operator $C$ contains the vector potential term. It is worth noting that vector potentials allow us to implement Hadamard coins using relativistic analogy. Relativistic limit of Hadamard DTQW was considered in \cite{kurzynski}, where the mass term in the corresponding Dirac equation violated Lorentz covariance. Before we plug Eqs. (\ref{pot}), (\ref{m}) and (\ref{vecpot}) into Eq. (\ref{genqw}), let us emphasize that in some cases it might be convenient to consider inverse tangent of both potentials, like we did previously with the mass, because of the periodicity of the arguments in the exponents. Moreover, since we consider division of positions into edges and vertices, scalar potential terms should act on vertices, whereas mass and vector potential terms should act on edges. As a result the double step operator can be taken as
\begin{equation}\label{double}
U^2=SCSV=e^{ip\sigma_z}e^{i\left(\tilde{m}_{j,j\pm 1}\sigma_y-\tilde{A}_{j,j\pm 1}\sigma_z\right)}
e^{ip\sigma_z}e^{i\tilde{u}_j},
\end{equation}
where $\tilde{A}_{j,j+1}=\arctan\left(\frac{1}{2}\frac{\text{Im}[H_{j,j+1}]}{\text{Re}[H_{j,j+1}]}\right)$, $\tilde{m}_{j,j+1}=\arctan\left(\frac{1}{2\text{Re}[H_{j,j+1}]}\right)$, and $\tilde{u}_j=\arctan(H_{jj})$.

A similar conversion can be done for higher-dimensional quantum walks. Moreover, it is clear that using the above analogy one can convert the corresponding DTQW back into CTQW. However, there are certain kinds of DTQW which cannot be easily converted into CTQW, therefore let us now concentrate on the reverse problem of conversion of DTQW into CTQW. An example of a case in which the conversion is problematic is a DTQW with position dependent coin, where not only the coin flip ratio $\theta_j$ depends on position, but also the generating operator $\sigma_j$ changes from position to position. In the previous section we considered DTQW version of a weakly coupled spin protocol and observed that the transfer can be suppressed due to a change of the first and the last coin flip generators from $\sigma_y$ to $\sigma_x$. The coin degree of freedom is a qubit, therefore the coin generator is a linear combination of $\sigma_x$, $\sigma_y$ and $\sigma_z$. In the Bloch sphere picture, coin operation is a rotation of a qubit about an axis $\vec{n}$, where the generator of rotation is $\vec{n}\cdot\vec{\sigma}$. In general, $\vec{n}$ can point in any direction on a surface of a three-dimensional sphere, however in the case of the Dirac equation (\ref{dirac}) it is bounded to the $yz$-plane. Due to this fact, simple conversion from DTQW into CTQW is not possible. As a result, DTQW offers a greater possibility of control over the transport than is allowed in the CTQW case.

Up to now we have considered only one-dimensional graphs, like chains. In two, and higher, dimensions one encounters even more complicated obstacles. Firstly, the typical formulation of a DTQW on a $d$-dimensional grid uses a $2d$-dimensional coin. Each coin dimension corresponds to a different direction for the walker to take in the next step. This is in contrast to the Dirac formulation of relativistic motion, where for example in the three-dimensional case the {\it coin} space is only four-dimensional. The conditional shift operator of the corresponding DTQW is of the form
\begin{equation}\label{multishift}
\text{Exp}\left(i\sum_j p_j\otimes \gamma_j\right),
\end{equation} 
where $j=1,2,\dots,d$ enumerates different orthogonal directions, $\gamma_j= |2j\rangle\langle 2j| - |2j-1\rangle\langle 2j-1|$ acts on the coin space and $p_j$ is the corresponding momentum operator for direction $j$. Note that $[\gamma_j,\gamma_k]=0$ for all values of $j$ and $k$, whereas the corresponding Dirac matrices anti-commute $\{\tilde{\gamma}_j,\tilde{\gamma}_k\}=2\delta_{jk}$. Moreover, $\tilde{\gamma}_j^2=\openone$ for all $j$, whereas for DTQW $\sum_j \gamma_j^2=\openone$.  Due to this reason, one is unable to recover the relativistic energy relation $E^2=|\vec{p}|^2+m^2$, but rather obtains $E^2=\sum_j |\alpha_j|^2 p_j^2+\beta^2$, where $\sum_j|\alpha_j|^2=1$ and $\beta^2$ is a parameter related to properties of the relevant coin operator, which can be interpreted as the mass. Since our method of conversion is closely related to the similarity of quantum walks to the Schr\"odinger and the Dirac equations, it is somehow expected that problems with conversion of DTQW into CTQW appear when the description of the DTQW departs from the Dirac one.

It is important to mention here the research on quantum walk search protocols on $d$-dimensional grids. The efficiencies of the two protocols, the DTQW \cite{shenvi, ambainis} and the CTQW \cite{childs4}, are quite similar. Only in dimensions $d=2,3$ the DTQW protocol is faster than the corresponding CTQW protocol. In \cite{childs3}, the authors of the CTQW protocol improved the efficiency of their model, however this was done at the cost of an additional {\it coin} degree of freedom, which had to be included --- departure from the standard CTQW structure was inevitable. The heart of the DTQW protocol \cite{shenvi} is the so called Grover coin, which for $d=2$ is of the following form
\begin{equation}\label{grover}  
C_G=\frac{1}{2}\begin{pmatrix} -1 & 1 & 1 & 1 \\ 1 & -1 &1 & 1 \\ 1 & 1 & -1 & 1 \\ 1 & 1 & 1 & -1 \end{pmatrix}.
\end{equation}
For the two-dimensional Grover walk the probability distribution remains localized at the initial position for all time. This effect was explained by Inui {\it et. al.} \cite{inui}, who noticed that eigenvalues of the Grover walk are highly degenerated. More than half of the eigenvalues are $\pm 1$. There is no CTQW analogy of the Grover walk. Since the Grover walk obeys translational symmetry, the corresponding CTQW version should also possess this property. However, for a translational symmetric CTQW on a $N\times N$ grid with periodic boundary conditions, coupling constant $J=|J|e^{i\varphi}$ and diagonal term $A$, the set of eigenvalues is given by $A+2|J|\left(\cos(\varphi - 2\pi k_x/N)+\cos(\varphi - 2\pi k_y/N)\right)$, where $k_x,k_y=0,1,\dots,N-1$, and it is clear that there is no possibility of such a high degeneracy.

\section{Conclusions}

In this work we studied the DTQW version of the quantum state transfer, which is originally described as a continuous time process. While in the continuous time scenario the transfer properties depend on couplings between neighboring positions, in the discrete time case it is the coin operator which is responsible for the perfect transmission from one position to another. We applied the analogy between the DTQW and the Dirac equation and between the CTQW and the Schr\"odinger equation to show that both coins and couplings can be interpreted as the mass of a particle, which allowed us to transform tje CTQW into the DTQW. We examined the DTQW versions of perfect state transfer protocols studied in \cite{christandl,wojcik1,wojcik2}. We found that in DTQW versions not only the particle, but also its intrinsic coin state is perfectly transfered. The general method of the CTQW transformation into the DTQW has been also discussed. Finally, we argued that some DTQW's do not have the corresponding CTQW versions due to the fact that in some cases the dynamics of DTQW is much richer than the CTQW one.

One of the biggest advantages of CTQW is that it can be easily defined on any graph, whereas DTQW is natural on regular graphs only, where the same coin degree of freedom can be used for all vertices. However, many problems in physics, as well as in computer science, are defined on regular graphs. The above discussion shows that in many cases DTQW is more general than CTQW, since it allows to observe some effects which cannot be observed in CTQW. It would be interesting to find phenomena which can be observed in CTQW only.

\section{Acknowledgements}

PK would like to thank Jiannis Pachos for stimulating discussions and Ravishankar Ramanathan for help with preparation of this manuscript. This work is supported by the National Research Foundation and Ministry of Education in Singapore.


\begin{thebibliography}{99}


\bibitem{graphene}
A. K. Geim and K. S. Novoselov,
Nature Materials {\bf 6}, 183 (2007).

\bibitem{ol}
D. Jaksch, C. Bruder, J.I. Cirac, C.W. Gardiner, P. Zoller,
Phys. Rev. Lett. {\bf 81} 3108 (1998).

\bibitem{linopt}
 E. Knill, R. Laflamme, and G. J. Milburn,
 Nature {\bf 409}, 46 (2001). 

\bibitem{hetero}
C. Delerue and M. Lannoo, 
{\it Nanostructures: Theory and Modelling}, Springer (2007).

\bibitem{childs1}
A. M. Childs, 
Phys. Rev. Lett. {\bf 102}, 180501 (2009).

\bibitem{lovett}
N. B. Lovett, S. Cooper, M. Everitt, M. Trevers, and V. Kendon,
 Phys. Rev. A {\bf 81}, 042330 (2010).

\bibitem{FG}
E. Farhi and S. Gutmann,
Phys. Rev. A {\bf 58}, 915 (1998)

\bibitem{Ahar}
Y. Aharonov, L. Davidovich, and N. Zagury, 
Phys. Rev. A, {\bf 48}, 1687, (1993).

\bibitem{Bose}
S. Bose, 
Contemp. Phys. {\bf 48}, 13 (2007).

\bibitem{Kay}
A. Kay,
Int. J. Quantum Inf. {\bf 8}, 641 (2010) 

\bibitem{christandl}
M. Christandl, N. Datta, A. Ekert, and A. J. Landahl
Phys. Rev. Lett. {\bf 92}, 187902 (2004) 

\bibitem{strauch1}
F. W. Strauch,
 Phys. Rev. A {\bf 74}, 030301 (2006) 

\bibitem{childs2}
A. M. Childs,
Comm. Math. Phys. {\bf 294}, 581 (2010) 

\bibitem{strauch2}
F.W. Strauch, 
Phys. Rev. A {\bf 73}, 054302 (2006).

\bibitem{bracken}
A.J. Bracken, D. Ellinas, I. Smyrnakis, 
Phys. Rev. A {\bf 75}, 022322 (2007).

\bibitem{kurzynski}
P. Kurzy\'nski,
Phys. Lett. A {\bf 372}, 6125 (2008).

\bibitem{wojcik1}
A. W\'ojcik, T. {\L}uczak, P. Kurzy\'nski, A. Grudka, T. Gdala, and M. Bednarska, 
 Phys. Rev. A {\bf 72}, 034303 (2005). 

\bibitem{wojcik2}
A. W\'ojcik, T. {\L}uczak, P. Kurzy\'nski, A. Grudka, T. Gdala, and M. Bednarska, 
Phys. Rev. A {\bf 75}, 022330 (2007).

\bibitem{wojcik3}
A. W\'ojcik, T. {\L}uczak, P. Kurzy\'nski, A. Grudka, M. Bednarska, 
Phys. Rev. Lett. {\bf 93}, 180601 (2004).

\bibitem{childs3}
A. M. Childs, J. Goldstone,
Phys. Rev. A {\bf 70}, 042312 (2004).

\bibitem{shenvi}
N. Shenvi, J. Kempe, K. Whaley,
Phys. Rev. A {\bf 67}, 052307 (2003)

\bibitem{ambainis}
A. Ambainis, J. Kempe, and A. Rivosh, 
Proc. 16th ACM-SIAM SODA, 1099 (2005)

\bibitem{childs4}
A. M. Childs, J. Goldstone,
Phys. Rev. A {\bf 70}, 022314 (2004)

\bibitem{inui}
N. Inui, Y. Konishi, N. Konno,
Phys. Rev. A {\bf 69}, 052323. (2004). 

\end{thebibliography}
\end{document}